\documentstyle[epsfig]{article}
\def\be{\begin{equation}}
\def\ee{\end{equation}}
\def\bea{\begin{eqnarray}}
\def\eea{\end{eqnarray}}

\begin{document}

\begin{flushleft}
{\large \bf Comment on "Reverse Brazil Nut Problem:
Competition between Percolation and Condensation"
}
\end{flushleft}

In a recent paper \cite{hql01} the Brazil nut problem (BNP), where 
particles with large diameters rise to the top when exposed to vertical
shaking and the reverse Brasil nut problem (RBNP) where the large
particles segregate to the bottom were investigated. In realistic
situations these effects are driven or at least accompanied by global 
convection. Convection rolls are observed already in mono-disperse systems:
appearantly vibrations lead to temperature gradients with temperature
defined through the average kinetic energy of the particles.
In order to suppress convection the authors present Molecular Dynamics
(MD) simulations for binary hard sphere mixtures at
{\em fixed\/} temperature. It appears legitimate to first
consider this simplified situation although it might be rather
remote from reality.
Depending on the mass and diameter ratios of the two
hard sphere species, the simulations show pronounced BNP and RBNP
behaviours. For an explanation a new {\em condensation versus percolation\/}
driven segregation mechanism was put forward, which has attracted
much attention in the press \cite{01}. The purpose of this comment 
is to show that the proposed mechanism is incorrect. All the
effects observed in the MD simulations can be understood by
simple conventional thermodynamics.

\vspace{0.4cm}

Because in the solid phase particles may no longer exchange
positions \cite{h99}, we conjecture that demixing must happen 
already in the fluid phase. Therefore
we consider a binary hard sphere mixture with species $A$ and
$B$ having masses $m_A > m_B$ and diameters $d_A > d_B$ in
the liquid phase with an appropriate equation of state (EOS),
e.g. the Mansoori-Carnahan-Starling-Leland
equation \cite{mcsl71}, with $\xi_i = (n_A/n) \, d_A^{\, i} 
+ (n_B/n) \, d_B^{\, i}$
\bea
P  =  n \, T \cdot Z(x_A,x_B) \, ,   
&& Z(x_A,x_B)  =  \frac{1}{1-x} +  
\frac{3\,x}{(1-x)^2} \, \frac{\xi_1 \xi_2}{\xi_3} +
\frac{x^2 \, (3-x)}{(1-x)^3} \, \frac{\xi_2^3}{\xi_3^2} \, ,
\eea
where $n=n_A + n_B$ is the particle density and
$x = x_A + x_B = (\pi /6) d_A^{\, 3} n_A + (\pi /6) d_B^{\, 3} n_B$
the packing fraction. It is then straightforward to compute the
Helmholtz free energy \cite{ysl00} under gravity
\bea
&& F = E - T\cdot S \, .
\eea
Variation of $F$ with respect to the particle densities $n_A$ and
$n_B$ for fixed particle numbers $N_A$ and $N_B$ leads to
two coupled nonlinear equations which are solved numerically.
In the limit $N_A d_A^{\, 3} \ll N_B d_B^{\, 3}$ {\em and\/} $d_A \gg d_B$
these equations respect Archimedes' principle. Note however, that the
initial condition of equal layer numbers $N_A d_A^{\, 2} = N_B d_B^{\, 2}$
used in \cite{hql01} is incompatible with this limit.
Consequently Archimedes' principle need not be obeyed in the
presented MD simulations in contrast to our na\"\i ve expectation.
Calculated filling fractions $x_A(z)$ and
$x_B(z)$ as functions of the height $z$ are depicted in Fig. 1 for various
mass ratios $m_A/m_B$ keeping the diameter ratio $d_A/d_B=2$
and particle number ratio $N_A/N_B=1/4$ fixed.
\begin{figure}
\begin{center}
\epsfig{figure=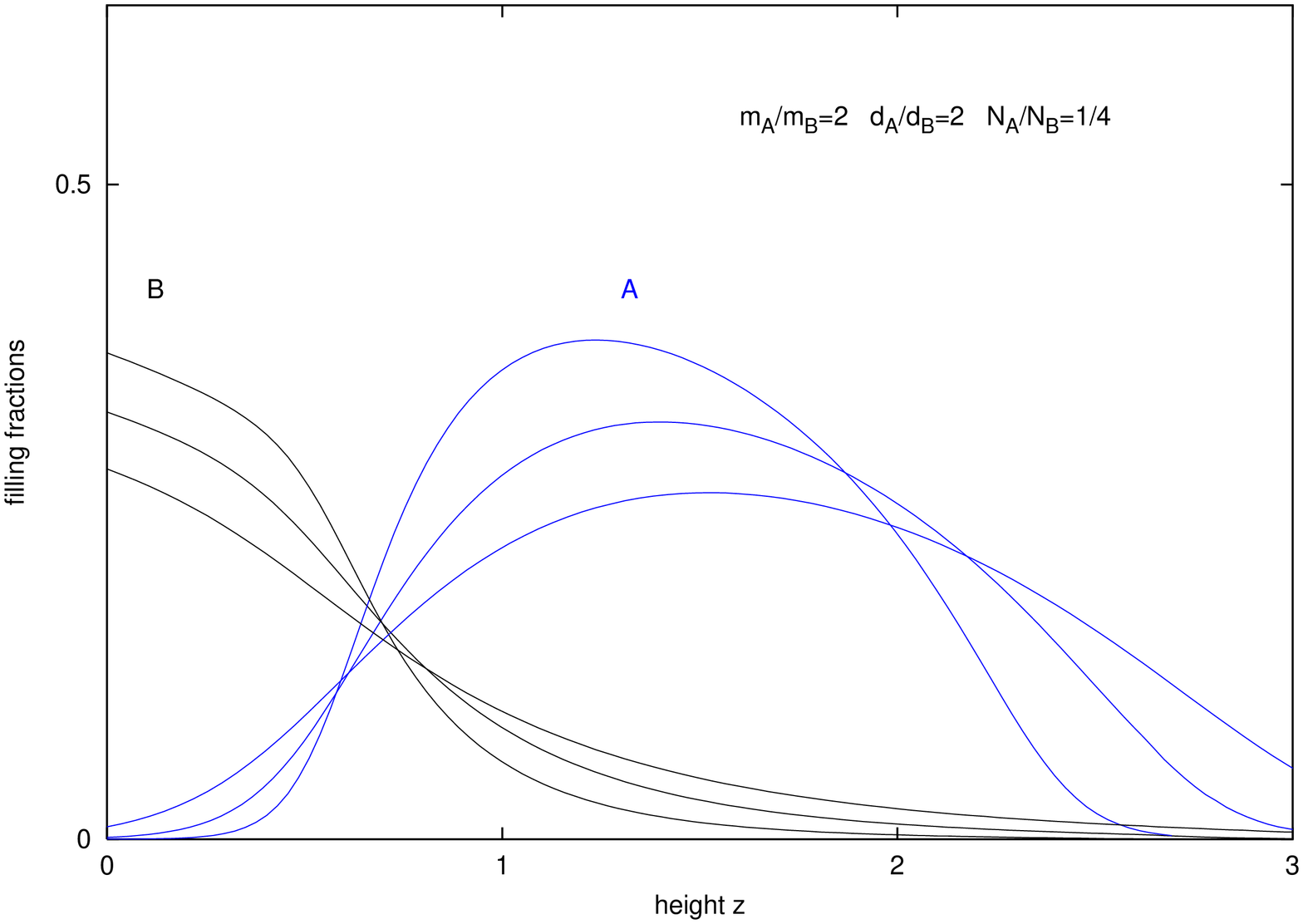,width=6cm,angle=270}
\epsfig{figure=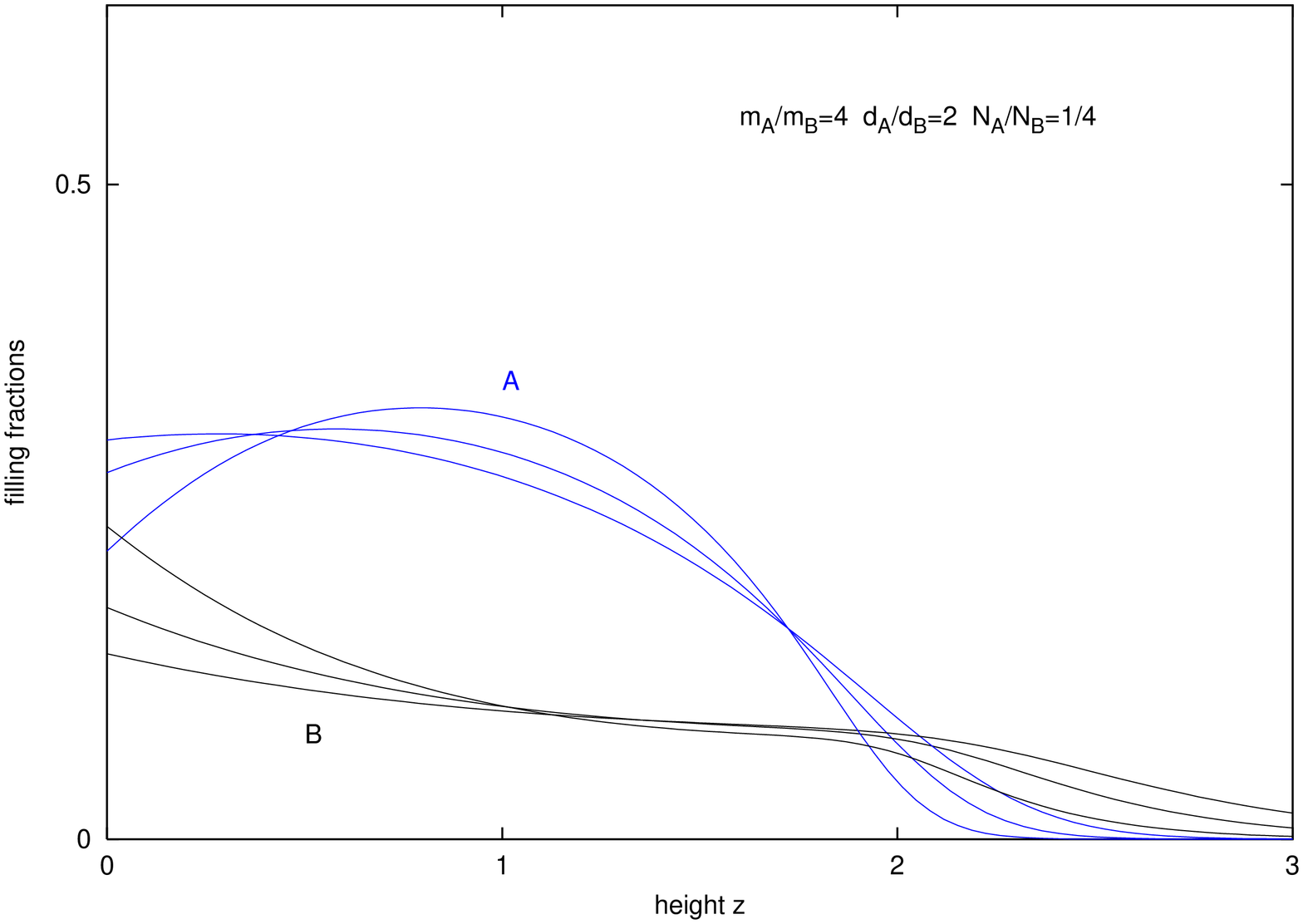,width=6cm,angle=270}
\epsfig{figure=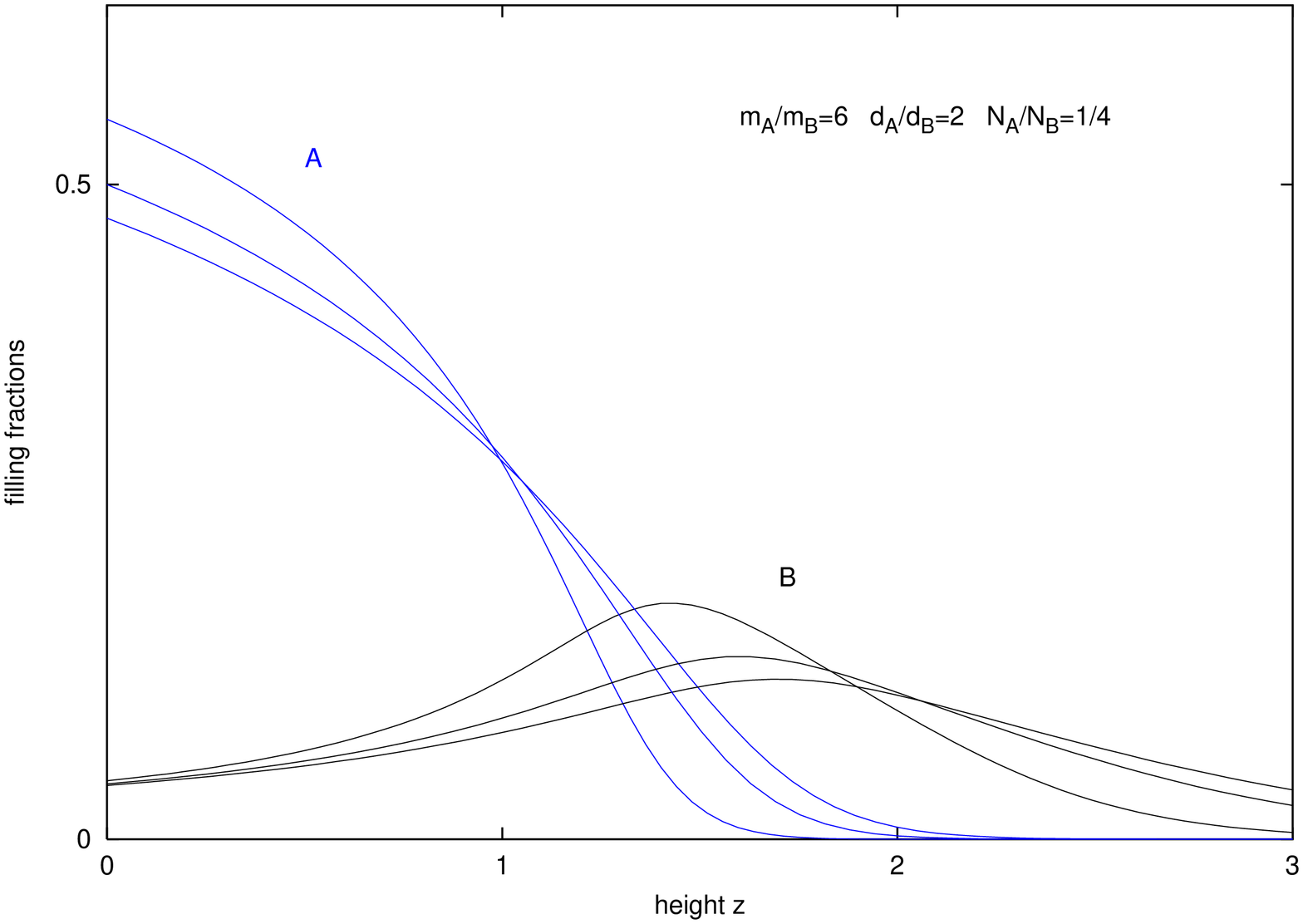,width=6cm,angle=270}
\protect\caption[]{Calculated filling fractions $x_A$ and $x_B$
of binary hard sphere liquids under gravity for
various mass ratios. 
The length unit equals the container size used in the
MD simulations of Ref. \cite{hql01}. 
The BNP and RBNP
are clearly recognized. With decreasing temperature
the profiles are shifted to smaller hights $z$.
}
\end{center}
\end{figure}
The density profiles compare very well with the MD simulations
presented in \cite{hql01}. For $m_A/m_B=2$ we encounter
the BNP with almost pure B-liquid at the bottom.
The $m_A/m_B=6$ case corresponds to the RBNP with an
incomplete demixing at the bottom. For $m_A/m_B=4$
the two liquids remain mixed. All these details agree
with the corresponding MD simulations. But note that
in our calculation the temperatures $T>T_C \geq T_A$
lie above the temperature $T_C$ where the binary liquid
begins to solidify: the total packing fraction lies 
sufficiently below close packing everywhere.
In contrast the temperatures used in \cite{hql01} are
supposed to be quenched between $T_B<T<T_A$ 
($T_A$ and $T_B$ represent the temperatures where condensation
sets in for the separated species \cite{h99}, see below)
such that parts of the sample are "frozen".

We must conclude that demixing indeed occurs in the
liquid phase. It is clearly due to the competition between 
{\em gravity and entropy\/} rather than to that of 
{\em condensation and percolation\/}.
There is nothing mysterious or special going on and all the effects
are understood by conventional thermodynamics. Finally,
if the temperature is further decreased, the binary liquid 
eventually begins to condensate locally according to its 
filling fractions $x_A$ and $x_B$ in the fluid phase. 
In the RBNP this starts from the bottom whereas in the BNP,
depending on the parameters, species $A$
in the middle may condensate first (cf. Fig. 1).
However, there exists no condensation driven demixing.

\vspace{0.4cm}

In order to discuss the transition from the BNP to the RBNP
and also to understand the effects on an even simpler basis,
we may consider the two species separated: $A$ above $B$ for the BNP
and $B$ above $A$ for the RBNP. The advantage is that we may then
use the much simpler EOS for one-component hard 
sphere fluids e.g. the Carnahan-Starling expression \cite{cs61}
\bea
& P = n \, T \, Z(x) \,  , \qquad &
Z(x) = \frac{1+x+x^2-x^3}{(1-x)^3}
\eea
obtained from (1) by setting $d_A = d_B$. The details of the EOS
are unimportant here, in fact, every $Z(x)$ which grows sufficiently
fast with $x$ (eigen-volume effect) leads to similar results.
We assume that at a filling fraction 
$x_0 \simeq 0.637$ equal to {\em random\/} 
close packing \cite{ttd00}, instead of FCC or HCP close packing 
$x_0=\pi / \sqrt{18} \simeq 0.74048$, condensation sets in, which
leads to temperatures
\bea
T_A = \frac{\pi}{6} \, \frac{N_A m_A d_A^{\, 3}}{x_0 Z(x_0)} 
\, \frac{g}{\mbox{area}} \, , \qquad
&&  T_B = \frac{\pi}{6} \, \frac{N_B m_B d_B^{\, 3}}{x_0 Z(x_0)} 
\, \frac{g}{\mbox{area}}
\eea
(gravity $g$) which exceed those in \cite{h99} by
roughly a factor of $3$. 
We choose $x_A(0)=x_0$ at the bottom of the RBNP which
corresponds to the temperature
$T_{RBNP} = T_A + (d_A/d_B)^3 \, T_B > T_A + T_B$ 
where this system is just still fluid.
Then also the BNP configuration is 
fluid, because there "freezing" sets in at lower temperatures
(either at $T_{BNP}= T_A$ at the boundary of the two species
or at $T_{BNP} = (d_B/d_A)^3 \, T_A + T_B$ at the bottom, 
depending on the parameters used). With the above choice
the filling fractions at the bottom and at the boundaries
are fixed by the corresponding
pressures and the Helmholtz free energies $F_{BNP}$ and
$F_{RBNP}$ may be computed. For
$F_{BNP} < F_{RBNP}$ the free energy favors the BNP and
vice versa. We consider the curves $F_{BNP} = F_{RBNP}$
which separate BNP from RBNP for the following cases:
\begin{itemize}
\item[(i)] equal number of particles $N_A=N_B$
\item[(ii)] equal number of layers $N_A d_A^{\, 2} =N_B d_B^{\, 2}$
\item[(iii)] equal volumes $N_A d_A^{\, 3} =N_B d_B^{\, 3}$
\item[(iv)] equal masses $N_A m_A = N_B m_B$.
\end{itemize}
The last case is not explicitly shown in Fig.2 because 
it almost coincides with case (ii). The MD simulation
results, case (ii), are also shown with circles (BNP),
boxes (RBNP) and triangles (mixed). In the vicinity
of the curves mixing is expected, demixing occcurs
for parameter combinations lying sufficiently far away. We find that the
particle numbers are of minor influence. This is expected:
the BNP and RBNP should not be
easily reversed by adding or subtracting a few particles
of either species.
\begin{figure}
\begin{center}
\epsfig{figure=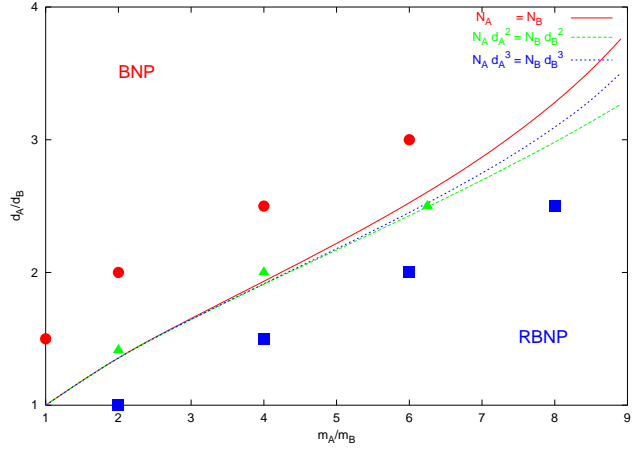,width=6cm,angle=270}
\protect\caption[]{Curves which separate BNP from
RBNP for different particle number ratios in the
mass ratio and diameter ratio parameter space.
The MD simulation results of Ref. \cite{hql01}
are also shown.
}
\end{center}
\end{figure}

Lets contrast our results  with the "percolation-condensation" formula
\bea
&& \frac{T_A}{T_B} = \frac{N_A m_A d_A^{\, 3}}{N_B m_B d_B^{\, 3}}
= \frac{d_A^{\, 3}}{d_B^{\, 3}}
\eea
of Ref. \cite{hql01}. This formula
seems to give reasonable results
$d_A/d_B = (m_A/m_B)^{1/2}$ for the equal--number--of--layers case (ii),
but certainly a straight line would not be worse.
For equal volumes, case (iii), (5) leads to
$d_A/d_B = (m_A/m_B)^{1/3}$ and the $m_A/m_B=6, \, d_A/d_B=2$ 
system is predicted BNP, which is incorrect. 
The calculated profiles look very similar to the corresponding
ones of case (ii)
(Fig. 1) with the only difference that $x_B$ is somewhat larger
because of the relatively larger particle number $N_B$. 
An MD calculation certainly would confirm this. For equal particle
numbers, case (i), the BNP is obtained for $m_A<m_B$ wheras
the equal mass case (vi) is predicted to be always mixed.
All these implications are wrong and we must conclude that
the formula (5) works only approximately for case (ii) by pure accident.

\vspace{0.4cm}

Preparing this comment, I realized that J.A. Both together
with one of the authors of Ref.\cite{hql01}, D.C. Hong, 
have just submitted a paper 
to the archive \cite{bh01}, which treats the problem along
similar lines. This paper, however, lacks the
clear statement that the "condensation - percolation theory" 
is incorrect. In view of  
the publicity given to this picture 
\cite{01} such a statement would seem appropriate.

\bigskip
\begin{flushleft}
H. Walliser \\[5mm]
{\small \it 
Fachbereich Physik, Universit\"at Siegen,  D-57068 Siegen, Germany} 
\end{flushleft}

\end{document}